\documentclass{article}

\usepackage{amssymb}
\usepackage{amsmath}
\usepackage{supertabular}
\usepackage{longtable}
\usepackage{array}
\usepackage{amstext}
\usepackage{epsfig}
\begin{document}
\title{Scaling solutions on a brane}
\author{N. Yu. Savchenko and  A. V. Toporensky}
\date{}
\maketitle
\hspace{-6mm}
{\em Sternberg
Astronomical Institute, Moscow University, Moscow 119899, Russia}

\begin{abstract}
We investigate the dynamics of
a flat isotropic brane Universe with two-component matter source:
perfect fluid with the equation of state $p=(\gamma-1) \rho$
and a scalar field with a power-law potential $V \sim \phi^{\alpha}$.
The index $\alpha$ can be either positive or negative. We describe
solutions for which the scalar field energy density scales as a
power-law of the scale factor (so called scaling solutions).
In the nonstandard brane regime when the brane is driven by energy density
square term these solutions 
are rather different from their analogs in the standard cosmology.
A particular
attention is paid to the inverse square potential. Its dynamical
properties in the nonstandard brane regime
are in some sense analogous to those of the
exponential potential in the standard cosmology. Stability analysis
of the scaling solutions are provided. We also describe solutions
existing in regions of the parameter space where the scaling solutions
are unstable or do not exist.
\end{abstract}

\section{Introduction}

Cosmological evolution of a brane world has become a matter of intense
investigation after the famous paper of L. Randall and R. Sundrum
\cite{R-S}. Due to additional terms, which appear in effective
Einstein equations on a brane, the brane cosmology has some
important differences
from the standard cosmological scenario.

The field equations on the brane are~\cite{oldMaeda,Maartens}
\begin{equation}
G_{\mu \nu}=\kappa^2 T_{\mu \nu} +
\tilde \kappa^4 S_{\mu \nu} - {\cal E}_{\mu \nu}.
\end{equation}
Here $\tilde \kappa=m_5^{-3/2}$ is the fundamental 5-dimensional gravitational constant,
$\kappa=m_4^{-1}$ is the effective 4-dimensional gravitation constant on the brane.
Bulk corrections to the Einstein equations on the brane are of two forms:
there are quadratic energy-momentum corrections via the tensor $S_{\mu \nu}$
and nonlocal effects from the free gravitational field in the bulk
${\cal E}_{\mu\nu}$.
 The nonlocal
corrections can be decomposed into a scalar,
vector and tensor parts~\cite{Maartens}.
On FRW brane there is only a scalar part ${\cal U}$ which
decays during the expansion of the Universe as ${\cal U} \sim a^{-4}$ where
$a$ is the scale factor. This means that this term behaves exactly as
a radiation fluid, and for this reason it is often called as "dark
radiation" despite its pure geometrical nature. It should be noted, however,
that "dark radiation" can be either positive or negative, and in the latter
case it gives rise to some cosmological solutions, impossible in the
standard cosmology (for example, the collapse of a flat brane Universe
\cite{Santos}).

More important differences between standard and brane cosmologies
appear at early stage of the Universe evolution, where quadratic
corrections proportional to $S_{\mu\nu}$ dominate over
the standard terms proportional to $T_{\mu\nu}$. The rate of expansion
of the Universe changes with respect to matter density, this leads to
interesting peculiarities of the brane scenario.

The case of matter source in the form of perfect fluid with
the equation of state $p=(\gamma-1)\rho$ have been extensively
investigated during last three years, see, for example,
\cite{C-S1, C-S2}.

 The next step is to consider a scalar field
on a brane. Even in the standard cosmology the dynamics of
Universe filled by a scalar field is much more complicated in
comparison to the perfect fluid case \cite{Star}.
The effective $\gamma$ of the field
can vary in the whole possible interval $\gamma \in (0, 2)$. This
property, if the scalar field potential is not too steep,
leads to existence of inflation and, in the case of
positive spatial curvature, to the possibility of bouncing solutions
and dynamical chaos.
However, if the potential is steep enough, the situation changes
and this transition occurs for potentials as steep as exponential:
$V(\phi) \sim \exp{\lambda \phi}$ with large enough $\lambda$ can not
provide inflation \cite{Liddle} and bounces \cite{Topor}. It already
have been noticed that in brane cosmology we have a similar picture
but with inverse square potential $V(\phi)=B \phi^{-2}$ playing the
role of exponential potentials in the standard cosmology \cite{our}.

In the present paper we consider a brane Universe filled by
a scalar field {\it and} a perfect fluid. In the ordinary cosmology
exponential potentials in this situation have again 
a particular property -- 
in this case a tracker solution (the energy density of a scalar
field tracks exactly the energy density of an ordinary matter)
exists \cite{Wands,Liddle2}. Not surprising that a tracker
solution on a brane have been found for the inverse square potential.
It was done by K.-I. Maeda \cite{Maeda1} in the particular case
of ordinary matter in the form of radiation fluid ($\gamma=4/3$).
We present a general form of brane tracker solution for an
arbitrary $\gamma$. We also study other power-law potentials
which give us brane scaling solutions -- solutions with scalar
field energy density scales as some power of a scale factor,
when the perfect fluid, being the dominant component, has an
energy density  scaling as a possibly different power
(we use the terminology of \cite{Liddle2}). Such
potentials are often used in modern quintessence models \cite{Maeda1,Mizuno,
Sami,Sahni} to explain observable acceleration of our Universe.

Our analysis can be useful for understanding  evolution of a quintessence
field before the radiation-dominated stage, for example, during inflation.
Though for the full treatment of this problem it is necessary to consider
a more complicated case of a two-field dynamics (quintessence and
inflaton fields), we can use the approximation of a perfect fluid
if $\gamma$ of the matter varies slowly enough.

In the present paper we consider only a flat isotropic brane.
Shear and curvature terms can modify the scaling solutions, in the
standard cosmology it was shown in \cite{Coley1, Coley2}. A two-component
(a scalar field and a perfect fluid) brane dynamics with nonzero shear
and curvature has already become a matter of investigations \cite{Hoogen1,
Hoogen2}. We leave the problem of scaling solutions on an anisotropic
brane to a future work.

\section{Tracker solutions}
\subsection{Basic equations}
We start with system of equations describing a scalar field
on a flat isotropic brane \cite{Varun}. There are the Raychaudhuri equation
\begin{equation}
\dot H=-H^2-\frac{\kappa^2}{6}(\rho + 3p) -
\frac{\kappa^2 \rho}{6 \lambda}(2\rho +3p)-\frac{2{\cal U}}
{\kappa^2 \lambda},
\end{equation}
the Friedmann equation
\begin{equation}
H^2=\kappa^2\frac{\rho}{3}+\kappa^2\frac{\rho^2}{6 \lambda}
+\frac{2 {\cal U}}{\kappa^2 \lambda},
\end{equation}
the Klein-Gordon equation for the scalar field
\begin{equation}
\ddot \phi + 3H\dot \phi + V'(\phi)=0,
\end{equation}
and the evolution equation for the "dark radiation"
\begin{equation}
\dot {\cal U}+\frac43 H {\cal U} =0.
\end{equation}
Here $\lambda$ is the brane tension.
The overall pressure and energy density in (2) -- (3) are the sums
$$
p=p_m+p_{\phi}, \qquad \qquad \rho=\rho_m+\rho_{\phi},
$$
where the perfect fluid equation of state $p_m=(\gamma_m-1) \rho_m$,
$\gamma_m={\rm const}$  and the scalar field equation of state
 $p_{\phi}=(\gamma_{\phi}-1) \rho_{\phi}$ with $\gamma_{\phi}$  changing with time.
The pressure $p_{\phi}$ and the energy density $\rho_{\phi}$ of the scalar field
are given by
$$
p_{\phi}=\frac{\dot \phi^2}{2}-V(\phi), \qquad
\rho_{\phi}=\frac{\dot \phi^2}{2}+V(\phi).
$$

Using Eq. (4) we can write the dynamical equation for $\gamma_{\phi}$ as
\begin{equation}
\dot \gamma_{\phi} =3H\gamma_{\phi}(\gamma_{\phi}-2)
-2\gamma_{\phi}\frac{V'}{\dot \phi}.
\end{equation}
In this section we study the particular potential
$V(\phi)=B\phi^{-2}$, for this case the Eq.(6) becomes
\begin{equation}
\dot \gamma_{\phi}=3H\gamma_{\phi}(\gamma_{\phi}-2)+
\frac{\sqrt{2\gamma_{\phi}}}{\sqrt{B}}\rho_{\phi}(2-\gamma_{\phi})^{3/2}.
\end{equation}

We would like to notice an interesting difference between scalar
field dynamics in ordinary cosmology and on a brane. In standard
cosmology a constant coefficient in field potential can be absorbed
by rescaling scale factor and time without changing a general
dynamical picture. In brane cosmology it is not the case, and coefficient
$B$ plays an important role, as we will see later.

After introducing the deceleration parameter $q$ as
$$
\dot H=-(1+q) H^2,
$$
dimensionless variables
$$
\Omega_{\rho m}=\frac{\kappa^2 \rho_m}{3H^2}, \quad \Omega_{\rho \phi}=
\frac{\kappa^2 \rho_{\phi}}{3H^2}, \quad \Omega_{\lambda m}=
\frac{\kappa^2\rho_m^2}
{6\lambda H^2}, \quad \Omega_{\lambda \phi}=\frac{\kappa^2\rho_{\phi}^2}
{6\lambda H^2}, \quad U=\frac{2{\cal U}}{\lambda
\kappa^2 H^2}
$$
and a new time variable $\tau$ such that $\frac{dt}{d\tau}=\frac{1}{H}$
we can rewrite our system in the form
\begin{multline}
q=\frac12\Omega_{\rho m}(3\gamma_m-2)+\frac12\Omega_{\rho \phi}
(3\gamma_{\phi}-2)
+\Omega_{\lambda m}(3\gamma_m-1)+{}\\
\Omega_{\lambda
\phi}(3\gamma_{\phi}-1)+
\sqrt{\Omega_{\lambda m}\Omega_{\lambda \phi}}(3(\gamma_m+\gamma_{\phi})-2),
\end{multline}
\begin{equation}
1=\Omega_{\rho m}+\Omega_{\rho \phi}+ U+
\Omega_{\lambda m}+\Omega_{\lambda \phi}+2\sqrt{\Omega_{\lambda m}
\Omega_{\lambda \phi}},
\end{equation}
\begin{equation}
\Omega_{\rho m}'=\Omega_{\rho m} [2(1+q)-3\gamma_m], \quad
\Omega_{\rho \phi}'=\Omega_{\rho \phi} [2(1+q)-3\gamma_{\phi}],
\end{equation}
\begin{equation}
\Omega_{\lambda m}'=\Omega_{\lambda m} [2(1+q)-6\gamma_m], \quad
\Omega_{\lambda \phi}'=\Omega_{\lambda \phi} [2(1+q)-6\gamma_{\phi}],
\end{equation}
\begin{equation}
U'=2 U (q-1),
\end{equation}
\begin{equation}
\gamma_{\phi}'=3\gamma_{\phi}(\gamma_{\phi}-2)+
\frac{\sqrt{72\gamma_{\phi}\Omega_{\lambda \phi}}}{\sqrt{B}}
m_5^3(2-\gamma_{\phi})^{3/2}.
\end{equation}
Here the prime denoted derivative with respect to $\tau$.

Note, that in the case of several matter sources nonlinear
crossing terms appear in the r.h.s. of Eqs. (8) -- (9).

In this paper we study only a pure nonstandard regime when
the contribution from linear terms proportional to $T_{\mu \nu}$
in (1) is negligible in comparison to
the quadratic energy-momentum terms proportional to
$S_{\mu \nu}$. Therefore
we put $\Omega_{\rho m}$
and $\Omega_{\rho \phi}$ to zero. For the potential $V(\phi)=B \phi^{-2}$
the Hubble parameter $H$ decouples (in standard regime this property
belongs to an exponential potential \cite{Dunsby}) and we have
a compact phase space unless a "dark energy" density $U < 0$.

\subsection{Equilibrium points}
Now we list equilibrium points for the system (8) -- (13)
which can be stable in future direction with
corresponding results of their stability analysis. To simplify
the formulae we use the dimensionless coefficient $b \equiv B/m_5^6$.
Remember, that $\Omega_{\rho m}$ and $\Omega_{\rho \phi}$ are set to zero.
As for the full system (8) -- (13) the equilibrium points obtained
correspond to an intermediate brane behavior, when some asymptotic
regime have already been established, but the brane is still driven
by the quadratic terms in (1). First, consider the case of $U = 0$.

\begin {description}
\item[I]
$$
\gamma_{\phi} = \frac{2}{1+b/8}, \qquad \Omega_{\lambda m} =0,
\qquad \Omega_{\lambda {\phi}}=1.
$$
Stable for $\gamma_m > \frac{2}{1+b/8}$.

\item[II]
$$
\gamma_{\phi} = \gamma_m, \quad \Omega_{\lambda \phi}=\frac{b \gamma_m}
{8(2-\gamma_m)}, \quad \Omega_{\lambda m}=1+\frac{b \gamma_m}
{8(2-\gamma_m)}-\frac{\sqrt{b\gamma_m}}{\sqrt{2(2-\gamma_m)}}.
$$
Exists and stable for $\gamma_m < \frac{2}{1+b/8}$.
\end{description}

The point {\bf I} is the regime of scalar field dominance. We see
that depending on $b$ the scalar field density can mimic an
ordinary matter in the whole interval of possible values
$\gamma_{\phi} \in (0, 2)$. Naturally, this point is stable
for $\gamma_m > \gamma_{\phi}$.
The point {\bf II} represents the brane tracker solution with
$\Omega_{\lambda \phi}/\Omega_{\lambda m}={\rm const}$, which was
obtained in \cite{Maeda} for the particular case of $\gamma_m = 4/3$.

The case of $U > 0$ leads to the following changes. The point {\bf I}
disappears for $b>16$, the point {\bf II} becomes unstable for
$\gamma_m > 2/3$. In these regions a new stable point appears:
\begin{description}
\item[III]
$$
\gamma_{\phi}=2/3, \qquad \Omega_{\lambda \phi}=\frac{b}{16}, \qquad
U = 1-\frac{b}{16}, \qquad \Omega_{\lambda m}=0.
$$
Exists for $b<16$, stable for $\gamma_{\phi} > 2/3$.
\end{description}
This is an analog of tracker solution when the "dark energy"
plays the role of matter. The ratio $\Omega_{\lambda \phi}/ U$
remains constant. Of course, for $2/3 < \gamma_m < 4/3$, this situation
("dark energy" dominates the ordinary matter) changes after
the low-energy stage begins and $\Omega_{\rho m}$ becomes important.
Zones of stability for the future direction
in the plane $(b, \gamma_m)$ for the case of
$U > 0$ are plotted in Fig.1.

\begin{figure}
\epsfxsize=5cm
\centerline{\epsfbox{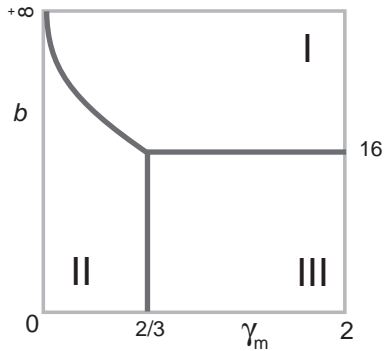}}
\caption{Stability zones of the equilibrium points for the case $U
>0$ on the plane $(b, \gamma_m)$ where $b \equiv B/m_5^3$.}
\end{figure}

If $U$ is negative, the phase space becomes incompact. Physically
this means that a recollapse of the brane Universe is possible. Our
numerical studies indicate that in zones of stability of points {\bf I}
and {\bf II} the recollapse occurs for large enough initial $|U|$, while
for small $|U|$ the Universe tends asymptotically to
corresponding stable points {\bf I} or {\bf II} with $U=0$. On the contrary,
in the region {\bf III} there are no stable equilibrium points, corresponding
to initial negative $U$. We have confirmed numerically, that all trajectories
from the region $b<16, \gamma_m>2/3, U<0$ recollaps. It is worth
to remind, however, that all this results are obtained for a pure
nonstandard regime. When the full system of equations (8) -- (13) is
considered, some trajectories may still not recollaps until
transition to the standard regime occurs, and, if $2/3 < \gamma_m < 4/3$
the recollapse will never occur.

\section{General power-law potentials}
\subsection{Scaling solutions}

In this section we describe solutions with scalar field energy density
scales exactly as some power (in general, different from the fluids one)
of the scale factor. The case of scalar field
dominance have been described in \cite{Maeda}, therefore we concentrate
our effort to the fluid dominance case. Our work generalize \cite{Maeda}
to fluids with $\gamma \ne 4/3$.

It is worth to note that these solutions are not future attractor in the
sense of the previous section even
if we do not consider the transition to the standard regime. Indeed,
the dynamical evolution may break down the fluid dominance.
Nevertheless, it is an interesting example of transient behavior.
This behavior is followed by a regime of a scalar field dominance.
Late-time scalar field dominance in the standard regime leads to
inflation \cite{Liddle2} and provides a possible explanation
of accelerated expansion of our present Universe.

We will follow the method of \cite{Liddle2}, where a similar class of
solutions have been found in the standard cosmology. Suppose that the scalar
field energy density behaves as $\rho_{\phi} \sim a^{-n}$, where
$a$ is the scale factor. Since the Klein-Gordon equation is not modified
on a brane, this property, as in the standard case, requires that
the ratio of a scalar field kinetic energy density and a total scalar field
energy density remains constant \cite{Liddle2}:
\begin{equation}
\frac{\dot \phi^2/2}{\rho_{\phi}}=\frac{n}{6}.
\end{equation}

The perfect fluid with equation of state $p=(\gamma-1)\rho$
has the energy density $\rho \sim a^{-m}$ where $m=3 \gamma$. In
the pure nonstandard brane regime the fluid dominance leads to
\begin{equation}
a \sim t^{1/m}.
\end{equation}
All our considerations remain valid also for the case of
the "dark energy" dominance (in this case, of course, ${\cal U}>0$),
because the "dark energy" behaves as a "normal" matter with $m=4$.
Due to (15), Eq.(4) becomes
\begin{equation}
\ddot \phi=-\frac{6}{m}\frac{1}{t}\dot \phi \frac{dV}{d\phi}.
\end{equation}
Eq. (14) gives us
$$
\dot \phi \sim t^{-n/2m}.
$$
Integrating this equation we obtain
\begin{equation}
\phi=A t^{1-n/2m}.
\end{equation}
Substituting (17) into (16) and solving for $V(\phi)$ we find the
potentials which allow the scaling behavior in the nonstandard
brane regime
\begin{equation}
V(\phi)=A^{2-\alpha} \frac{2 (3(\alpha-2)-\alpha m)}{\alpha m (\alpha-2)^2}
\phi^{\alpha},
\end{equation}
where
\begin{equation}
\alpha=\frac{2n}{n-2m}.
\end{equation}
It means that, as in the standard cosmology, scaling solutions exist
for power-law potentials $V(\phi) \sim \phi^{\alpha}$. If $\alpha$
is given, the scalar field energy density is proportional
to $a^{-n}$, where
\begin{equation}
n=\frac{2 \alpha}{\alpha-2} m.
\end{equation}

Though this expression differs only for the factor $2$ in the numerator
from its analog in the standard cosmology \cite{Liddle2}, the properties
of brane scaling solutions are rather different. Let us begin with the
case of negative $\alpha$. In the standard scenario the
energy density of a scalar field with $\alpha<0$ always falls slower
than the density of a perfect fluid while in the brane cosmology
it falls slower for $\alpha>-2$ and faster for $\alpha<-2$. Second,
the positivity of the potential in (18) requires that
$\alpha (3-m) <6$, which indicates that for $m>3$ (a matter with a positive
pressure) and
\begin{equation}
\alpha< \frac{6}{3-m}
\end{equation}
the scaling solution does not exist. On the contrary, in the standard
cosmology the scaling solution exists for an arbitrary negative $\alpha$
independently on $m$.

Consider now the case of positive $\alpha$. The positivity of (18) leads
to $\alpha (3-m)>6$. It means that for $m>3$ scaling solutions do
not exist for any $\alpha$. In particular, they do not exist for
the case of $\gamma=4/3$, investigated in \cite{Maeda}. On the other
side, for $m<3$ the situation is similar to the standard cosmology~--
scaling solution exists for sufficiently steep potential with $\alpha$
satisfying in the brane case
\begin{equation}
\alpha>\frac{6}{3-m}.
\end{equation}

\subsection{Stability analysis}

We have to analyze the stability of the scaling solutions (17)
with a power-law potential (18). To do this we use new variables
\begin{equation}
\label{scalnewvar}
\tau=e^t\!,\qquad u(\tau)=\frac{\varphi(\tau)}{\varphi_0(\tau)}\,,
\end{equation}
where $\varphi_0(\tau)$ is the exact solution given by (17)
Then the equation for scalar field becomes
\begin{equation}
\label{scalfield}
u''+u'\left(\frac{2+\alpha}{2-\alpha}+\frac 3m\right)+\frac 2{\alpha-2}
\left(\frac{\alpha}{\alpha-2}-\frac
3m\right)\left(u-u^{\alpha-1}\right),
\end{equation}
where the prime denotes the derivative with respect to $\tau$. This
second-order differential equation can be divided into the autonomous
system
\begin{equation}
\label{scalsys}
\begin{array}{l}
u'=p,\\[2mm]
p'=\frac 2{\alpha-2}\left(\frac 3m-\frac{\alpha}{\alpha-2}\right)
\left(u-u^{\alpha-1}\right)-\left(\frac{2+\alpha}{2-\alpha}+\frac
3m\right)p.
\end{array}
\end{equation}

Scaling solution corresponds to a critical point $(u,p)=(1,0)$.
Linearizing (\ref{scalsys}) about this point we find the eigenvalues of
these coupled equations
\begin{equation}
\label{scaleigen}
\lambda_{1,2}=\frac{\alpha+2}{2(\alpha-2)}-\frac 3{2m}\pm
\sqrt{\left(\frac{\alpha+2}{2(\alpha-2)}-\frac 3{2m}\right)^{\!2}
+\frac{2\alpha}{\alpha-2}-\frac 6m}.
\end{equation}
Then the condition for stability is the negativity of the real part of
both eigenvalues.

For $\alpha<0$ after analysis we find that scaling solution is a stable
attractor for all region of existence (21). And for
$\alpha>0$ there is an additional inequality
\begin{equation}
\label{scalineq}
\alpha>\frac{2m+6}{3-m}\,,
\end{equation}
which excludes a narrow domain from (22). For example,
for $V \sim \phi^4$ the scaling solution is stable if $m<1$, or,
equivalently, $\gamma_m<1/3$, this is exactly the condition for
a brane inflation driven by an ordinary matter. For $V \sim \phi^6$
we have $m<3/2$ (or $\gamma_m<1/2$).

\begin{figure}
\epsfxsize=10cm
\centerline{\epsfbox{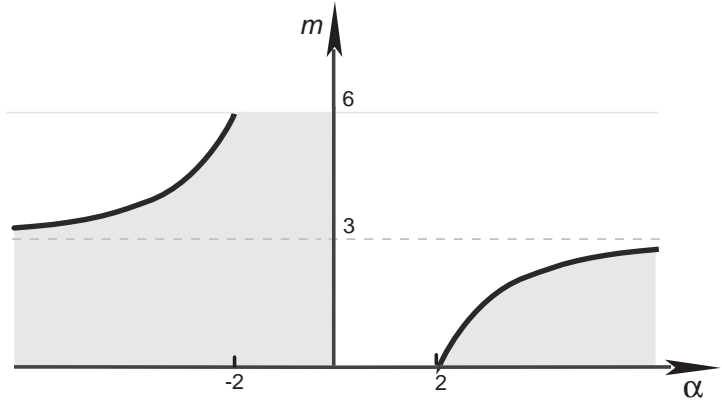}}
\caption{Stability zones of the scaling solutions (dark).
In the white zone with negative $\alpha$ the kinetic-term dominated
solution is stable, in white zone with positive $\alpha$ the scalar
field infinitely oscillates.}
\end{figure}

Zones of stability of the scaling solutions on the plane $(\alpha, m)$
are plotted in Fig.2.

\subsection{Other solutions}
In the end of this section we briefly describe solutions, existing
in the regions of the plane $(\alpha, m)$ where scaling solutions
do not exist.

 For positive $\alpha$ with even $n$ we have infinite damping oscillations
of the scalar field, similar to the standard case. Some asymptotic formulae
for potentials $V=\frac{1}{2}m^2 \phi^2$ and $\frac{1}{4}\lambda \phi^4$
have been given in \cite{Maeda}.

For negative $\alpha$ the other type of solution exists. In this kinetic-term
dominated solution the contribution of the potential for the energy
density of the scalar field is negligible in comparison to the
contribution of the kinetic term. Its explicit form is
\begin{equation}
\phi=\phi_0 \left(\frac{t}{t_0}\right)^{\frac{m-3}{m}},
\end{equation}
where $\phi_0$ and $t_0$ are integration constants. This solution
does not exist for $m \le 3$. In this case the scaling solution
is stable for all negative $\alpha$. For $m=4$ we recover the known
behavior $\phi \sim t^{1/4}$ \cite{Maeda}.

\section{Conclusions}
We have investigated a scalar field dynamics with  a power-law potential
on a brane filled by a perfect fluid with equation of state
$p_m=(\gamma_m-1)\rho_m$
in a regime when quadratic energy-momentum correction
dominates. The tracker solution when the ratio of scalar
field and fluid energy densities remains constant is found for the scalar field
potential $V(\phi)=B \phi^{-2}$. Two other future asymptotic
for this case are described and zones of their stability in the plane
$(B, \gamma_m)$ are found.

We also describe scalar field evolution with $V(\phi) \sim \phi^{\alpha}$
in a brane Universe, dominated by  perfect fluid energy density. We have
found scaling solution, analogous to known in the standard cosmology.
The main differences between standard and brane cosmology are:
\begin{itemize}
\item For negative $\alpha$ in the standard cosmology the scaling
solutions exist for arbitrary $\gamma_m$ and $\alpha$, whereas
in the brane cosmology they do not exist for
$\gamma_m>1$ and large enough $|\alpha|$, which do not satisfy (21).
\item For positive $\alpha$ in the standard cosmology the scaling
solutions exist for any $\gamma_m$ if $\alpha$ exceeds some value,
depending on $\gamma_m$. In the brane cosmology the scaling
solutions exist only for $\gamma_m<1$ and large enough $\alpha$,
satisfying (27). They are absent for $\gamma_m>1$.
\end{itemize}

It is also interesting, that in the case of negative $\alpha$ the scalar
field energy density may fall more rapidly or more slowly than the
fluid density, depending on the sign of $\alpha+2$, whereas in the standard
scaling solution with negative $\alpha$ the ratio of scalar field and
fluid energy densities always increases. This was discovered
in the particular case of radiation fluid
in \cite{Maeda1}, where it was argued that this fact may be important
for quintessence scenario. In our paper we confirm that this property
do not depend on the state equation of the perfect fluid.

\section*{Acknowledgments}

The work was supported by RFBR grants Ns. 00-15-96699
and 02-02-16817.

\end{document}